\documentclass[doublecol]{epl2}
\usepackage{graphicx}

\title{On the destruction of the hidden order in URu$_2$Si$_2$ by a strong magnetic field}

\author{Julien Levallois\inst{1} \and Kamran Behnia\inst{2} \and Jacques Flouquet\inst{3} \and Pascal Lejay\inst{4} \and Cyril Proust\inst{1} }
\shortauthor{J. Levallois \etal}

\institute{
  \inst{1} Laboratoire National des Champs Magn\'{e}tiques
Puls\'{e}s (UMR 5147 CNRS-UPS-INSA), Toulouse, France

  \inst{2} Laboratoire Photons et Mati\`{e}re (UPR5-CNRS),
ESPCI, 10 Rue de Vauquelin, 75231 Paris, France

  \inst{2} Institut Nanosciences et Cryog\'{e}nie, SPSMS/MDN,
CEA-Grenoble, 38054 Grenoble, France

  \inst{2} Institut N\'{e}el, CNRS/UJF, B.P. 166, 38 042
Grenoble Cedex 9, France }


\abstract{ We present a study of transport properties of the heavy
fermion URu$_2$Si$_2$ in pulsed magnetic field. The large Nernst
response of the hidden order state  is found to be suppressed when
the magnetic field exceeds 35~T. The combination of resistivity,
Hall and Nernst data outlines the reconstruction of the Fermi
surface in the temperature-field phase diagram. The zero-field
ground state is a compensated heavy-electron semi-metal, which is
destroyed by magnetic field through a cascade of field-induced
transitions. Above 40 T, URu$_2$Si$_2$ appears to be a polarized
heavy fermions metal with a large density of carriers whose
effective mass rapidly decreases with increasing magnetic
polarization.}

\begin{document}

\maketitle

\section{INTRODUCTION}
Discovered more than twenty years
ago\cite{Palstra85,Schlabitz86,Maple86}, the order which emerges
in URu$_2$Si$_2$ below $T_0$=17.5 K remains as enigmatic as ever.
This unidentified order parameter (proposed to be associated with
orbital degrees of freedom\cite{Santini94,Kiss05}, a spin density
wave\cite{Ikeda98, Mineev05}, orbital magnetism\cite{Chandra02} or
magnetic helicity\cite{Varma06}) is commonly called the hidden
order (HO) state. For a long time, it was associated with a tiny
ordered moment ($M_0\sim 0.02\mu_B$ per U atom) \cite{Broholm87},
which did not match the large anomalies detected in the
macroscopic properties at $T_0$. A consensus on the extrinsic
nature of this tiny moment is gradually emerging, since its size
decreases with improvement in sample quality\cite{Amitsuka07}. On
the other hand, a true antiferromagnetic (AF) ground state with a
large ordered moment (0.3~$\mu_B/ U $) and an identical wave
vector (0,0,1) emerges above a rather small critical pressure
$P_x$ of 0.5~GPa\cite{Amitsuka07,Hassinger08}. The first order
HO-AF boundary meets the HO and AF lines at a tricritical point at
$P_c = 1.2$~GPa. The tiny moment at zero pressure appears to be a
consequence of the difficulty to obtain samples in which local
stress is totally removed\cite{Amitsuka07,Villaume08}.

The study of transport properties has established a radical
reconstruction of the Fermi surface below $T_0$. An early study of
the Hall effect (using a simplified one-band model) had already
concluded that the carrier density in the hidden-order state is as
low as 0.03 carriers per formula unit (f.u.)\cite{Schoenes87}
implying that the carrier density is more than one order of
magnitude lower than in the unordered state and what is predicted
by band calculations. A large drop in carrier density in the HO
state provides a natural explanation for a drastic enhancement of
phonon thermal conductivity \cite{Behnia05,Sharma06}. Moreover,
the carrier concentration deduced from the Hall data is compatible
with the largest of the small frequencies detected in the  de
Haas- van Alphen (dHvA) and Shubnikov-de Haas (SdH)
measurements\cite{Bergemann97,Keller98,Ohkuni99} and the large
magnetoresistance observed in the new generation of ultraclean
crystals\cite{Kasahara07}. Thus, the state which emerges below
$T_0$ is a dilute liquid of heavy quasi-particles, with Fermi
surfaces occupying a fraction of the Brillouin zone much smaller
than what it is expected according to band calculations. Such a
context indicates a Fermi surface reconstruction based either on a
nesting scenario associated with a density wave instability
\cite{Wiebe07} or on a change in the lattice symmetry.

However, several puzzles face any scenario based on nesting. The
first problem is to identify a plausible nesting vector. In
neutron scattering measurements in the hidden order phase, except
the extrinsic tiny antiferromagnetic moment at Q$_{AF}$=(0,0,1),
two main inelastic magnetic responses of Q$_{0}$=(1,0,0) and
Q$_1$=(0.4,0,1) have been detected. One suggested candidate is the
incommensurate wave vector Q$_1$=(0.4,0,1) for a spin density at
$T_0$ \cite{Wiebe07}. Under pressure above $P_x\simeq$0.5~GPa in
the AF phase at low temperatures, the inelastic response at Q$_0$
has disappeared while the excitations at Q$_1$ persists and has
shifted to a higher energy than at $P=0$ \cite{Villaume08}.
However, a nesting wave vector along either Q$_1$ or Q$_{0}$ is
not obvious from the point of view of band calculations in the
paramagnetic regime \cite{Harima}. The second problem comes from
the behavior of transport and thermodynamic properties under
pressure. While a phase transition clearly occurs between the
hidden ordered state (stabilized below $T_0$ for pressures below
$P_c$) and the AF phase (stabilized below $T_{N}$ for pressures
above $P_c$)\cite{Motoyama03}, resistivity and specific heat
anomalies remain unaltered across $P_c$\cite{Hassinger08}. This
indicates that the ordering leads to a radical reconstruction of
the Fermi surface both below and above $P_c$. In other words, if
there is nesting, it seems to be indifferent to presence or
absence of the large AF ordered moment. Recent inelastic neutron
diffraction experiments under pressure lead to propose that the
same wave vector Q$_{0}$ characterize the hidden order and the AF
phases \cite{Villaume08}. At $T_0$ or $T_N$, the crystal structure
changes from body centered tetragonal to tetragonal, implying a
carrier density drop by a factor between 3 and 5, according to
recent band structure calculations \cite{Harima,Elgazzar08}.

The hidden order is destroyed by the application of a strong
magnetic field\cite{deVisser87}. Interestingly, the application of
magnetic field, which weakens the partial Fermi surface
gap\cite{vanDijk97}, leads to an enhancement of the gap magnetic
excitations at Q$_{0}$, while the one at Q$_{1}$ slowly decreases
\cite{Bourdarot03}. Studies of
transport\cite{Bakker93,Kim03,Oh07,Jo07} and thermodynamic
properties\cite{Jaime02,Harrison03} have shown that a cascade of
phase transitions occurs at low temperature when magnetic field
ranging from 35~T to 39~T is applied parallel to the easy
\emph{c}-axis, leading to the destruction of the hidden order.
Above 39~T, a polarized paramagnetic (PPM) metal is restored via
successive jumps of magnetization near 0.3~$\mu_B$
\cite{deVisser87, Date94, Inoue01,Harrison04}. Thus, the
metamagnetic transition, a recurrent feature of heavy fermion
physics, takes a particularly complicated twist in presence of the
hidden order.

In this paper, we report on a study of transverse
magnetoresistance (MR) and Hall effect in URu$_2$Si$_2$ in pulsed
magnetic field up to 55~T. We also present the first measurements
of the Nernst effect in pulsed magnetic fields in the hidden order
phase.  According to our results, the large Nernst response which
emerges below T$_{0}$\cite{Bel04} is suppressed above 35~T. Hence,
it is a property of the hidden-order state. Moreover, and in
agreement with previous studies, we find that the hidden order and
the high-field PPM state are separated by two other unidentified
ground states. Thus, three successive field-induced Fermi surface
reconstructions occur before the stabilization of the polarized
high-field state. The latter happens to be a metal with a high
density of carrier, in sharp contrast with the zero-field one: It
has a large Fermi surface and low electronic mobility.


\section{Experimental}

\indent Single crystals of URu$_2$Si$_2$ were prepared in a
three-arc furnace under a purified argon atmosphere and annealed
under UHV for one week at 1050 $^{\circ}$C. Different samples with
typical residual resistivity $\rho_0 \approx$ 7 $\mu\Omega cm$ and
typical dimensions (2$\times$1$\times$0.03)~mm$^3$, have been
measured. In the four-probe resistivity measurements, the current
was injected along the $\it{a}$-axis of the crystals and magnetic
fields was applied along $\it{c}$-axis. The resistivity of each
sample was measured at a given fixed temperature during the
magnetic field pulse using a lock-in amplifier working at 50~kHz.

A special set-up to measure the Nernst effect in presence of a
pulsed field was designed. Its technical details will be published
elsewhere\cite{Proust08}. The transverse Nernst voltage produced
by a DC thermal gradient, $\Delta T \approx$1~$K$ along the sample
was measured with a low-noise pre-amplifier. The thermal gradient
and the temperature of the sample were measured with two
Chromel/Constantan thermocouples before and after pulsing the
magnetic field. The field dependence of the thermal gradient (as
expected from the field dependence of thermal
conductivity\cite{Behnia05}) was neglected. In the temperature
range of investigation, heat conduction is dominated by the phonon
contribution. The available data up to 12T\cite{Behnia05} suggest
that at $T$=3~K, the field-induced change in thermal gradient
could modify the magnitude of the Nernst signal by 50 percent.
This would not alter any of our conclusions. In order to get the
required signal/noise ratio, in particular to reduce mechanical
vibrations, we have used a magnet whose maximum field is 36~T.

\section{Cascade of phase transitions}

\begin{figure}[t]
\centering
\includegraphics[scale=1.2]{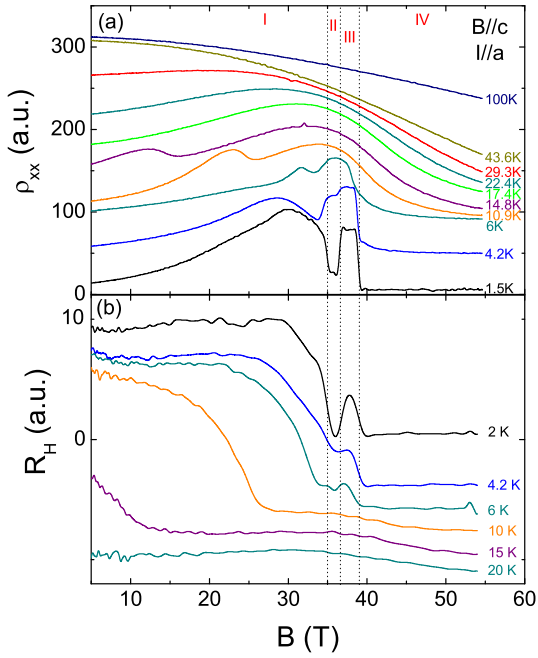}
\caption{(a) Magnetic field dependence of the transverse
magnetoresistance in URu$_2$Si$_2$ at different temperatures. The
labels on the left axis stand for the curve at T=1.5 K in
$\mu\Omega cm$. The other curves are shifted for clarity. The
labels I-IV denote the different parts of the phase diagram
\cite{Kim03} (b) Magnetic field dependence of the Hall coefficient
in URu$_2$Si$_2$ at different temperatures. The labels on the left
axis stand for the curve at T=2 K in $10^{-3} cm^3/C$. The other
curves are shifted for clarity.} \label{fig1}
\end{figure}

Fig.~\ref{fig1}(a) shows the field dependence of the transverse
magnetoresistance ($\textbf{I}\parallel\textbf{a}$ and
$\textbf{B}\parallel\textbf{c}$), above and below the hidden order
transition temperature $T_0$=17.5~K.

As originally reported by Bakker and co-workers\cite{Bakker93},
three field distinct field scales can be identified between 35~T
and 40~T. During the past few years, high-field studies of
specific heat\cite{Jaime02}, ultrasound \cite{Suslov03}
magnetization\cite{Harrison03}, resistivity\cite{Kim03,Jo07} and
Hall effect \cite{Oh07} have identified several phases (up to
five) in this field window. The three field scales identified in
our low field data allows to distinguish between four states,
which, following previous authors, are labelled from I to IV.

 \subsection{Phase I ($B<35~T$)}

\begin{figure}[t]
\centering
\includegraphics[scale=0.8]{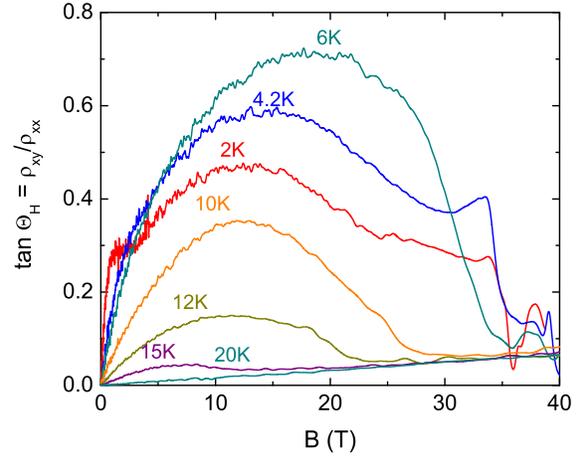}
\caption{Magnetic field dependence of the Hall angle of
URu$_2$Si$_2$ at different temperatures} \label{fig2}
\end{figure}

\begin{figure}[t]
\centering
\includegraphics[scale=0.8]{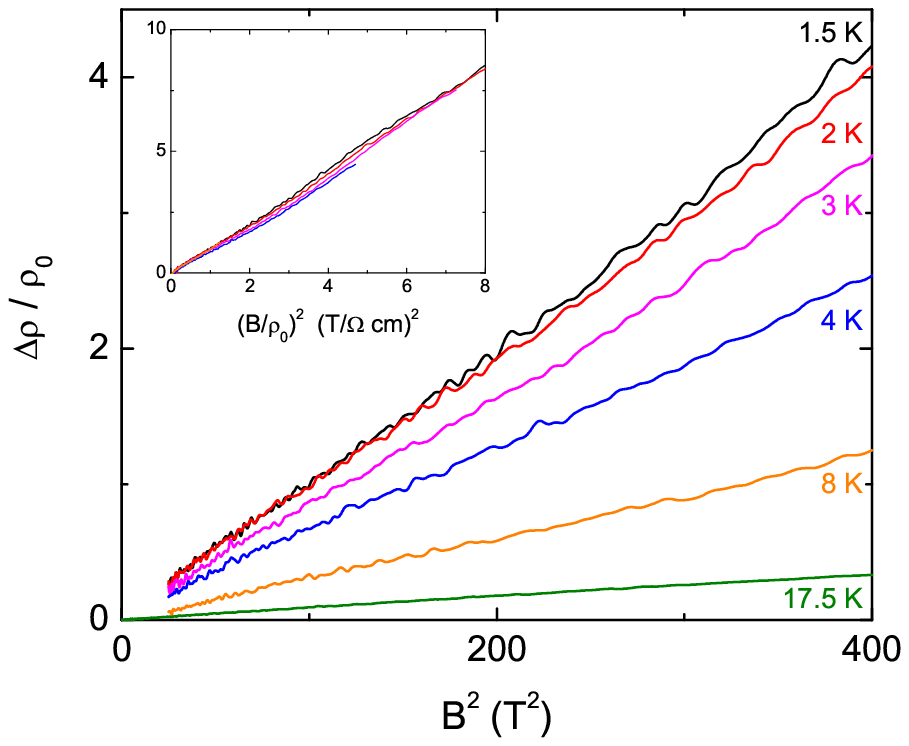}
\caption{Magnetoresistance plotted as B$^2$ for different
temperatures. The inset shows a Koehler plot of the
magnetoresistance at different temperatures.} \label{figMR}
\end{figure}

The zero-field ground state (i.e., the hidden order state) remains
the ground state up to $B_{HO}=35~T$. One of its remarkable
properties is a large Hall number ($R_H=\rho_{xy}/B$). Our data
yields $R_H\sim10^{-8}m^{3}/C$ at $T$=2~K. This magnitude is close
to the value reported in crystals of comparable residual
resistivity\cite{Bakker93,Bel04,Oh07}. In a simple one-band
picture, it would yield to an effective carrier density of
$n_H=1/eR_H\approx$0.05 carrier per U atom.

Band structure calculations \cite{Ohkuni99, Harima, Elgazzar08}
categorize URu$_2$Si$_2$ as a compensated metal. The fact that the
magnetoresistance does not saturate in the hidden order up to 20~T
in ultraclean crystals where $\omega_c\tau >> 1$ \cite{Kasahara07}
indicates that URu$_2$Si$_2$ is a compensated semi-metal.

Fig.~\ref{fig2} shows the magnetic field dependence of the Hall
angle defined as $\tan\Theta_H=\rho_{xy}/\rho_{xx}$. The entrance
in the hidden order state is concomitant with the appearance of a
large Hall angle. In a simple one-band picture the Hall angle is a
measure of $\omega_c\tau=\mu B$, where $\mu=e \tau /   m^*$ is the
mobility. Its large magnitude in the hidden order state is a
consequence of the enhanced carrier mobility.

In the hidden order state, the magnetoresistance is quadratic up
to $\approx$20~T(see fig.~\ref{figMR}). In a simple one-band
picture, the magnitude of the magnetoresistance is intimately
related to mobility $\mu$ via the relation
$\frac{\Delta\rho(B)}{\rho_{0}}\propto(\omega_c\tau)^2=(\mu B)^2$.
In our case, a field of 20 T leads to a 4-fold increase in
resistivity. It is instructive to compare the slope of
$\frac{\Delta\rho(B)}{\rho_{0}}$ as a linear function of $B^2$
with the one reported for an ultraclean crystal\cite{Kasahara07}.
The large difference (it was 3 T$^{-2}$ there compared to 0.011
T$^{-2}$ here) implies a mobility ratio equal to 16.5, in good
agreement with the 14-fold difference in the magnitude of residual
resistivities (7~$\mu\Omega$cm here and 0.5~$\mu\Omega$cm there).
This is a good check of the overall consistency of this analysis.
This is further confirmed by the validity of the Kohler's rule in
the hidden order state (see inset of fig.~\ref{figMR}) and by the
value of the mobility deduced from magnetoresistance,
$\mu\approx$0.1~T$^{-1}$, in good agreement with the low-field
slope of $\tan\Theta_H(B)$.

The combination of a small Fermi surface and high mobility suffice
to explain the large Nernst signal, which emerges in the
hidden-order state. \cite{Bel04} The Nernst coefficient is defined
as $N = E_y/\nabla_x T$, where $E_y$ is the electric field
generated by the combined application of a longitudinal thermal
gradient and a transverse magnetic field.

\begin{figure}[t]
\centering
\includegraphics[scale=0.8]{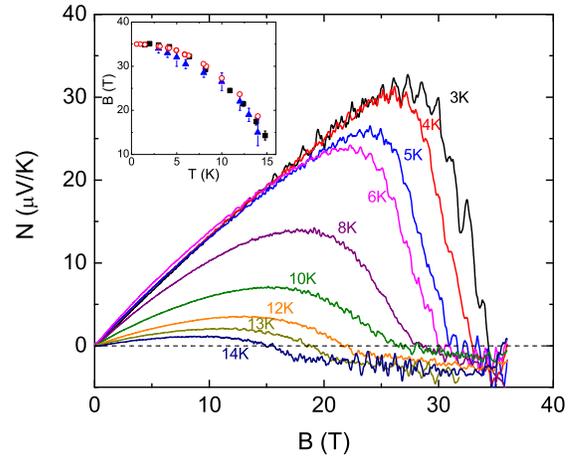}
\caption{Magnetic field dependence of the Nernst signal up to 36~T
at different temperatures. The curves have been smooth in order to
reduce the noise induced by mechanical vibrations seen at high
fields. The inset compares the phase diagram (T,B) of
URu$_2$Si$_2$ in the hidden order obtained from the present MR
(square) and Nernst effect (triangles) studies with steady field
measurements of the MR (open circles from \cite{Kim03,Oh07}).}
\label{fig3}
\end{figure}

In order to check that the large Nernst signal is a property of
the hidden order state, we have performed the first study of the
Nernst effect in pulsed magnetic fields up to 36~T.
Fig.~\ref{fig3} shows the field dependence of the Nernst signal at
different temperatures in the hidden order state. The magnitude of
the Nernst signal found here is twice lower than the data reported
for $B<$12 T\cite{Bel04}.  The discrepancy may be a consequence of
the uncertainty on the magnitude of thermal gradient. At low
temperature and for magnetic field up to $\sim$20~T, the Nernst
signal varies linearly with field and reaches a maximum of
$N\simeq$30~$\mu$V/K at $T$=3~K. Taking the sign change of the
Nernst signal changes as the signature of the destruction of the
hidden order state, we obtain the phase diagram $(B,~T)$ shown in
the inset of Fig.~\ref{fig3}, which is in agreement with the one
obtained by magnetoresistance measurements by following the
inflection point of the first drop in resistivity, both in pulsed
fields (squares) and steady fields (open circles) experiments. The
large Nernst signal is indeed a property of the hidden order
state.

A finite Nernst signal is expected in a multi-band metal with
different types of carriers\cite{Bel03}. Moreover, as illustrated
by the case of elemental bismuth \cite{Behnia07}, the Nernst
effect, which tracks the ratio of mobility to the Fermi energy
becomes particularly large in a clean semi-metal. In both
URu$_2$Si$_2$ and PrFe$_4$P$_{12}$ \cite{Pourret06}, the large
Nernst signal is a property of the ordered state, which can be
described as a semi-metal.

 \subsection{Phase II ($35~T < B <37~T$)}
Little is known about phase II, besides that the carrier density
is large. Indeed the Hall number is about 20 times lower than in
the hidden order state suggesting a large Fermi surface. The
reconstruction of the Fermi surface between phase I and II is
accompanied by a sharp drop in the mobility in the latter. The
small magnitude of the Nernst signal in this phase is, therefore
no surprise. Since both this phase and phase IV are high-density
metals, an open question is the distinction between these two
states. An appealing possibility is that a wave vector different
from $Q_0$ (maybe $Q_1$) could be a new wave vector leading to
less severe Fermi surface reconstruction.

\subsection{Phase III ($37~T < B <39~T$)}

At 37~T another Fermi surface reconstruction occurs. The Hall
number in phase III is rather large but four times lower than in
the hidden order state. It has a rather high resistivity which
does not vary much with field. All this suggest that it is not
simply the reentrant hidden order state. Moreover, since this
phase persists in $U(Ru_{0.96}Rh_{0.04})_{2}Si_2$ where the hidden
order phase collapse\cite{Oh07}, its presence appears to be
insensitive to the physical condition necessary for the emergence
of the hidden order. Field induced AF quadrupolar ordered phase
may occur, as observed for example in PrOs$_4$Sb$_{12}$.
\cite{Shiina04}.

\begin{figure}[t]
\centering
\includegraphics[scale=0.8]{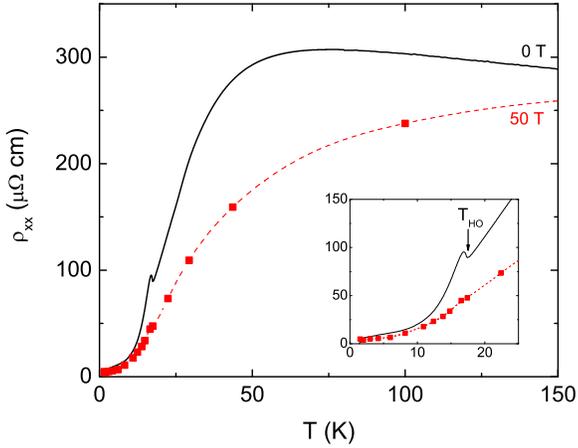}
\caption{Temperature dependence of the resistivity of
URu$_2$Si$_2$ in zero magnetic field and at B=50~T (red squares).
The inset shows a zoom at low temperature} \label{fig5}
\end{figure}

\subsection{Phase IV($B>39T$)}
The cascade of transitions ends in a final metamagnetic transition
at $B_M$=39~T associated with a magnetization jump of
0.3$\mu_B$\cite{Inoue01}, leading to a PPM at high fields. In
contrast with the hidden-order state, we will see later that the
PPM state is a metal with a much larger Fermi surface and much
lower mobility.

To sum up, phase I is a semi-metal which a) is compensated, b) is
a low carrier-density system, and c) has a rather high electronic
mobility. The cascade of phase transitions which take place
between 35~T and 39~T, induces several Fermi surface
reconstructions. Among all this complexity, URu$_2$Si$_2$ shares
at least one similarity with other heavy fermions compounds,
namely a polarized paramagnetic state reached above 39~T.

\section{Field-induced destruction of electronic correlations}
Fig.~\ref{fig5} compares the temperature dependence of resistivity
in the PPM state with the one in the HO, which are not so
different. In particular, as seen in the inset of Fig.~\ref{fig5},
both present a comparable resistivity below 10 K. However, in
spite of their superficial similarity, these two are quite
different metals. In the PPM state, the Hall coefficient is small,
$R_H$=0.47$\times$10$^{-9}$m$^3$/C, yielding a carrier density of
1.1 hole per U atom. Therefore, a residual resistivity of
$\rho_{xx}\approx$4.4~$\mu\Omega~cm$ implies a mobility of $\mu
\simeq$ 0.01 T$^{-1}$, which is one order of magnitude smaller
than the mobility found in the hidden order state.

\begin{figure}[t]
\centering
\includegraphics[scale=0.9]{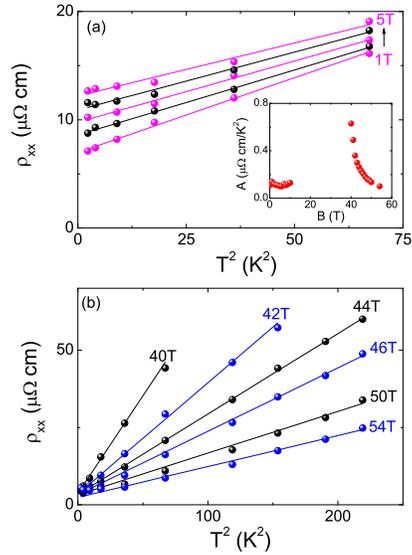}
\caption{ $\rho_{xx}$ plotted versus T$^2$ at different magnetic
fields (a) in the hidden order state, (b) in the PPM state. The
solid lines represent the fits to $\rho_0 + A T^2$ (see text).
Inset: Magnetic field dependence of the inelastic term A of the
resistivity.} \label{fig6}
\end{figure}

Let us turn our attention to the magnitude of the inelastic
resistivity and its field dependence. Fig.~\ref{fig6}(a) and
\ref{fig6}(b) show a comparison of the temperature dependence of
the transverse MR (plotted versus $T^2$) for different magnetic
fields, in the hidden order state and in the PPM state,
respectively. Due to the strong MR at low temperature and to the
anomalies of the MR around $B_{HO}$, it is not possible to
reliably extract the \emph{A} coefficient between 7~T and 40~T. As
seen in the inset of Fig.~\ref{fig6}(a), at 54 T, the \emph{A}
coefficient attains a magnitude comparable to its zero-field value
($\sim 0.15~\mu\Omega$cmK$^{-2}$). Using
$\gamma$=0.06~JK$^{-2}$mol$^{-1}$, the Kadowaki-Woods (KW) ratio
$A/\gamma^2$ at zero field can be estimated to be 40
$\mu\Omega$cmJ$^{-2}$mol$^{2}$K$^{2}$, four times larger than the
universal value introduced by Kadowaki and Woods\cite{KW86}.
According to available theoretical
treatments\cite{Hussey05,Kontani04}, the KW ratio is expected to
be enhanced when the carrier density decreases, a feature which
has been experimentally checked in other heavy-fermion semi-metals
such as PrFe$_{4}$P$_{12}$\cite{Pourret08}. Therefore, the
moderately enhanced KW ratio of the hidden order is in qualitative
agreement with its reduced carrier density.

\begin{figure}[t]
\centering
\includegraphics[scale=0.8]{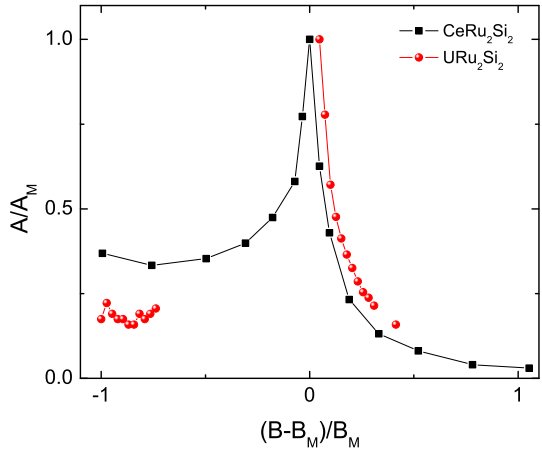}
\caption{Comparison of the magnetic field dependence of the
inelastic term of the resistivity, $A$, for URu$_2$Si$_2$ and
CeRu$_2$Si$_2$ . B$_M$ is the field which corresponds to the
metamagnetic transition.} \label{fig7}
\end{figure}


Regarding the high-field PPM state, in absence of specific heat
and/or dHvA data, the effective mass cannot be quantitatively
determined. Given its conventionally large carrier density, it is
natural to assume that the KW ratio is not enhanced in the PPM
state. Thus, the magnitude of A implies  $\gamma \simeq
0.1~JK^{-2}mol^{-1}$ at $B_M$. Obviously, the first order nature
of the metamagnetic transition in URu$_2$Si$_2$ does not wipe out
the observation of strong electronic fluctuations above $B=B_M$.
In agreement with this observation, Harrison \emph{et al.} have
suggested the presence of a putative quantum critical point at
$B=B_M$ in URu$_2$Si$_2$ \cite{Harrison03}. If the long range
ordering will collapse, such phenomena suggest the presence of a
metamagnetic critical point $B_M^*$ in the
temperature-pressure-magnetic field phase diagram. When the
ordering (either the hidden order or AF) is destabilized by an
external parameter (pressure and/or magnetic field), $B_M$ will
not collapse but ends up at a finite critical field value, at it
has been identified in the CeRu$_2$Si$_2$ familly
\cite{Flouquet06}. Another analogy between URu$_2$Si$_2$ and
CeRu$_2$Si$_2$ si the field dependence of the inelastic term of
the resistivity (see Fig~~\ref{fig7}), which appears to show a
critical behavior near $B_M$. However, for CeRu$_2$Si$_2$ even at
$P=0$, $B_M$ is a sharp crossover field between a low field state
dominated by AF fluctuations and a high field regime governed by
local fluctuations \cite{Flouquet02}. By contrast to
URu$_2$Si$_2$, the number of carrier estimated from Hall
measurements does not change significantly through $B_M$ in
CeRu$_2$Si$_2$\cite{Daou06, Kambe96}. A modification of the FS is
clearly detected at $B_M$ and scenarios based on the localization
of the 4\emph{f} cerium electrons versus a progressive evolution
of the Fermi surface where the heaviest sheet of the polarized
Fermi surface becomes completely filled and drops out of transport
and thermodynamic properties are still under discussion
\cite{Flouquet06, Kusminskiy07}. Let us remark that in the PPM
phase of URu$_2$Si$_2$ the derived large number of light carriers
near 1 hole per U atom is quite comparable to the value of holes
(O.5 hole per U atom) found in band structure calculations in
ThRu$_2$Si$_2$ \cite{Harima}. It corresponds to the limit of very
strong magnetic field where the U atoms are fully polarized and
decoupled from the light itinerant band, analogous to that of
ThRu$_2$Si$_2$.

In URu$_2$Si$_2$, the new ingredient is the interplay between the
ordered phases and Fermi surface reconstruction. Without this
reconstruction, the ground state of URu$_2$Si$_2$ would be a
non-ordered Kondo lattice with an intermediate valent character of
the uranium atoms, since the internal structure of the \emph{5f}
angular momentum is wiped out by the large damping of the
conduction electrons \cite{Villaume08, Janik08, Hassinger08b}.
URu$_2$Si$_2$ would therefore be a paramagnetic mixed valence
compound presenting a broad maxima of $C/T\simeq$300
mJK$^{-2}$mol$^{-1}$ at about $T\simeq$30~K and of the
susceptibility at $T\simeq$50~K \cite{Hassinger08b}. These maxima
are precursor of metamagnetic phenomena. Below, $T_0$, the
interplay between band structure and multi-ordering leads to
drastic change of spin dynamics. The decrease of the carrier
density reveals the local properties of the uranium atoms, which
is required for the establishment of multiple ordering. At $B_M$,
when the large number of carrier is recovered, $C/T$ reach a value
comparable to the maximum value in zero field above $T_0$. The
open question is the description of the uranium atoms in the
ordered and paramagnetic phases in terms of localization and
itineracy of the two 5f electrons (assuming that the uranium atom
is in the U$^{4+}$ configuration). It is worthwhile to compare the
$(B, T)$ phase diagram of URu$_2$Si$_2$ and the one of some mixed
valence systems, such as YbInCu$_4$\cite{Matsuda07} or doped Ce
metal \cite{Drymiotis05}, where an elliptic shape of the boundary
has also been reported. However, when discussing multipole
ordering, it should be noted that Yb$^{3+}$ and Ce$^{3+}$ are
Kramer's ions and their valence fluctuations involve non-magnetic
configuration. In the uranium case, the mixing will involve two
possible magnetic configurations: U$^{3+}$ and U$^{4+}$. Moreover,
fancy effects may occur due to crystal field in addition to the
transition at $T_0$, suspected to be non isostructural.

\section{CONCLUSION}
We have discussed the high field transport properties of
URu$_2$Si$_2$. The hidden order state can be described as a
semi-metal which is compensated, has a low carrier-density and has
a rather high electronic mobility. This naturally explains the
emergence of a large Nernst effect in this phase. Above the
destabilization of the hidden order state by a magnetic field of
35~T, a cascade of phase transition occurs and corresponds to
several Fermi surface reconstruction, whose origin must be
associated with different wave vectors. Above 39~T, a polarized
paramagnetic metal is recovered with a high density of carrier and
and low electronic mobility.

\acknowledgments We thank D. Aoki, W. Knafo, G. Knebel, and G.
Rikken for usefull discussions. Part of this work was supported by
the French ANR IceNET and EuroMagNET.

\end{document}